\documentclass[aip,reprint,citeautoscript,amsmath,amssymb,nofootinbib,longbibliography]{revtex4-2}

\usepackage{graphicx}
\usepackage{newtxtext}
\usepackage{newtxmath}
\usepackage{bm}
\usepackage{xcolor}
\usepackage[colorlinks=true,allcolors=blue!60!black]{hyperref}
\graphicspath{{./}}

\newcommand{\hw}{\hbar\omega_B}
\newcommand{\kz}{k_z}
\newcommand{\Ep}{E_\perp}

\makeatletter
\newenvironment{figurehere}{\def\@captype{figure}\setlength{\parindent}{0pt}}{}
\makeatother

\begin{document}

\title{Landau diamagnetism and the de Haas--van Alphen effect
from a single geometric construction}

\author{Sung-Hoon Lee}
\email{lsh@khu.ac.kr}
\affiliation{Department of Applied Physics, Kyung Hee University,
Yongin, Republic of Korea}

\date[]{}

\begin{abstract}
Landau diamagnetism is usually derived from the grand canonical
potential, a calculation that yields the correct susceptibility but
little physical insight, while the rule that extremal cross sections
govern the de Haas--van Alphen (dHvA) oscillations is commonly asserted
rather than exhibited. We present an
elementary zero-temperature construction in which the occupied states of
a free-electron gas are grouped, interval by interval along the field
direction, into the Landau levels onto which they condense. Within each
interval the field-induced energy cost reduces to a transfer of states
between two congruent triangles, and center-of-mass arithmetic yields
the Landau susceptibility. The construction fails only in a narrow neighborhood of an extremal
cross section of the Fermi surface; there the contribution oscillates
with the dHvA period, one oscillation for each Landau level that passes
through the extremal cross section.
Direct zero-temperature state counting locates the oscillation: the
residual is concentrated in the few intervals nearest the extremal cross
section, the summed contribution of the distant ones being negligible on
the scale of the peak oscillation amplitude.
\end{abstract}

\maketitle
\renewcommand{\thefootnote}{\ensuremath{\dagger}}

\section{Introduction}
\label{sec:intro}

\begin{figure*}[t]
\makebox[\textwidth][c]{\includegraphics{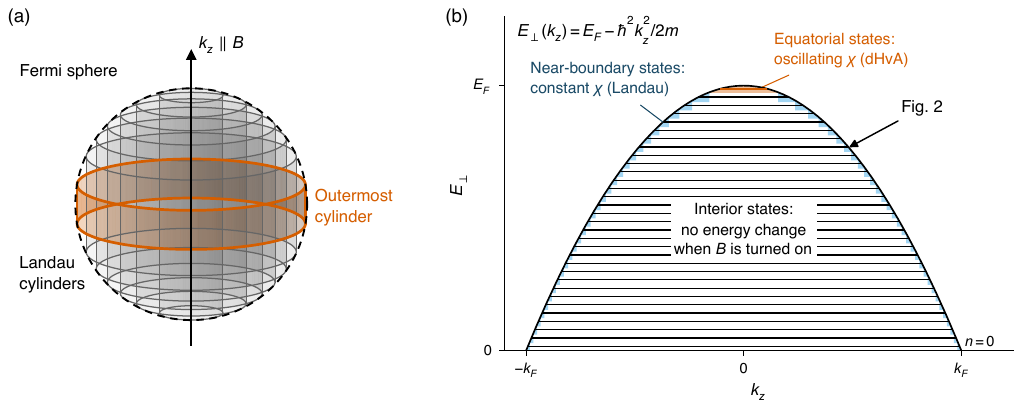}}
\caption{\label{fig:cylinders}%
(a)~Landau cylinders inside the Fermi sphere (dashed silhouette). The
outermost occupied cylinder (orange) is a thin band near the equator: its
level energy $E_n$ lies just below $E_F$, so only a short segment around
$\kz=0$ is occupied. (b)~The same system in the $(\kz,\Ep)$ plane, with a denser ladder (32 occupied levels; the spacing is still hugely exaggerated, $\hw/E_F\sim10^{-5}$ in reality). Landau cylinders are horizontal lines;
the occupied segments lie below the boundary parabola of
Eq.~\eqref{eq:boundary}. Each shaded triangle is the occupied part of one
strip $E_n\pm\tfrac12\hw$ within its tiling interval. On the flanks (blue)
the triangles tile the $\kz$ axis and each contributes the same constant
amount (Sec.~\ref{sec:triangle}). At the apex the boundary is flat and
never crosses the upper edge of the strip, so that one interval gives a
cap (orange) instead: a pair of slightly curved triangles back to back.
What sets this interval apart is that its contribution is not fixed. As
$B$ increases the topmost level rises through $E_F$, the cap collapses
and is renewed, and the magnetization oscillates (Sec.~\ref{sec:dhva}). Fully occupied interior strips regroup onto Landau levels, but their
net energy change cancels by symmetry. The
marked interval is magnified in Fig.~\ref{fig:triangle}.}
\end{figure*}

The orbital response of conduction electrons to a magnetic field presents a
curious situation. Its two central results are long established: the steady Landau diamagnetic susceptibility, $\chi_L = -\tfrac{1}{3}(m_0/m^*)^2\chi_P$ for an isotropic parabolic band, and the de Haas--van Alphen (dHvA) oscillations periodic in $1/B$ \cite{deHaas1930}. Yet the standard derivations stop short in complementary ways.

For the steady term the standard route is Landau's grand canonical
calculation with Euler--Maclaurin summation \cite{Landau1930,Peierls1955},
which produces the factor $-\tfrac13$ but offers no picture of
\emph{which} electrons respond or \emph{why} the energy rises; standard
references accordingly state the result and refer the derivation
elsewhere \cite{AshcroftMermin,Kittel8}. For the oscillations the
situation is complementary. The two-dimensional calculation, in which the
total energy develops a sawtooth as Landau levels empty, is elementary and
correct, but the passage to three dimensions is made by asserting that
orbits whose periods are stationary in $\kz$ dominate, the justification
being given as ``essentially\ldots\ a question of phase cancellation'' and
referred to more advanced treatments \cite{Kittel8}. The extremal-orbit
rule is thereby stated rather than exhibited. The rigorous result is of
course available: Lifshitz and Kosevich obtained the oscillatory
thermodynamic potential by stationary phase, and the dominance of extremal
cross sections follows from it \cite{LifshitzKosevich1956,Shoenberg}. What
the present construction adds is not rigor but an elementary account, in
closed form, of how the cost of the oscillation is assembled. The two-dimensional
mechanism, moreover, is parametrically suppressed in a three-dimensional
metal, where the Fermi energy is pinned by the overwhelming majority of
non-extremal states.

The Landau-tube construction \cite{AshcroftMermin} shows that the density
of levels at the Fermi energy is singular whenever a tube of extremal
cross section satisfies the quantization condition, but the argument stops
at the density of states: neither the magnetization nor any connection to
the steady term follows from it. Elementary derivations of the steady term
do exist. Pippard sketched a slice-by-slice picture in which each slice of
thickness $dk_z$ exchanges electrons with a reservoir held at the Fermi
energy, noted that the steady diamagnetism follows, and obtained the
oscillatory part by a Cornu-spiral summation over slices \cite{Pippard1960}.
Dupr\'e later gave an explicit $T=0$ derivation of the nonoscillatory
term by grouping the zero-field states into energy bunches and replacing
the values of a variable over the $\kz$ planes by a uniform average;
there the particle-number balance is established globally, after summing
over the planes, rather than within each interval. He noted that the
uniform approximation fails near the equator and gives a periodic term,
referring to Pippard for its analysis \cite{Dupre1981}. The same calculation is reproduced in
textbook form by Blundell \cite{Blundell} and has since been adapted to
electrons gyrating in a crystal lattice \cite{Olszewski2009}; the relation
between the Pauli and Landau responses has been revisited in a
methodological note by Batyev \cite{Batyev2009}. More recently Nikolaev
formulated the steady response of the free-electron gas and of monovalent
metals as a Fermi-surface effect, using a magnetic-tube construction that
localizes it to a narrow shell at the Fermi surface; for the free-electron
gas he also treated the oscillation, assigning it to irregular tubes in
the equatorial region \cite{Nikolaev2018}.

In this paper we give a single geometric construction that addresses these gaps by elementary state counting. The occupied states of the free-electron gas
at $T=0$ are grouped, interval by interval along $\kz$, into the Landau
levels onto which they condense (Secs.~\ref{sec:setup} and~\ref{sec:triangle}). Within one
interval, the entire energy cost of switching on the field reduces to the
transfer of states between two \emph{congruent triangles} in the plane
of $\kz$ versus the perpendicular energy $\Ep$; center-of-mass
arithmetic then gives an energy cost of $\tfrac{1}{24}\,p\,\hw$ per
$\kz$ point, with $p$ the Landau-level degeneracy, and summing over the
intervals, which tile the $\kz$ axis, yields the Landau susceptibility
(Sec.~\ref{sec:triangle}). The construction rests on a
single assumption: that the occupation boundary can be linearized across
one interval. That assumption fails only where the boundary is flat, i.e.,
near an extremal cross section, and the contribution of that neighborhood
oscillates: each period marks the passage of one
Landau level through the extremal cross section, with the Onsager period
$\Delta(1/B)=2\pi e/\hbar S_{\rm ext}$ set by the extremal cross-sectional area $S_{\rm ext}$ ($S_{\rm ext}=S_0=\pi k_F^2$ for the Fermi sphere; Sec.~\ref{sec:dhva}). Direct $T=0$ state counting makes the localization quantitative: of the
$43\,191$ bins resolved at $B\approx1\;$T, each combining the
symmetry-related intervals at $\pm\kz$, the extremal bin and its first
neighbor reproduce the oscillatory residual to within $1.2\%$ of its peak
amplitude (Supplementary Information). The relation to the treatments of Pippard and Dupr\'e is made precise
in Sec.~\ref{sec:discussion} and Appendix~\ref{app:dupre}, and to
Nikolaev's magnetic-tube analysis in Appendix~\ref{app:nikolaev}.

\section{Landau levels and Landau cylinders}
\label{sec:setup}

\begin{figure*}[t]
\makebox[\textwidth][c]{\includegraphics{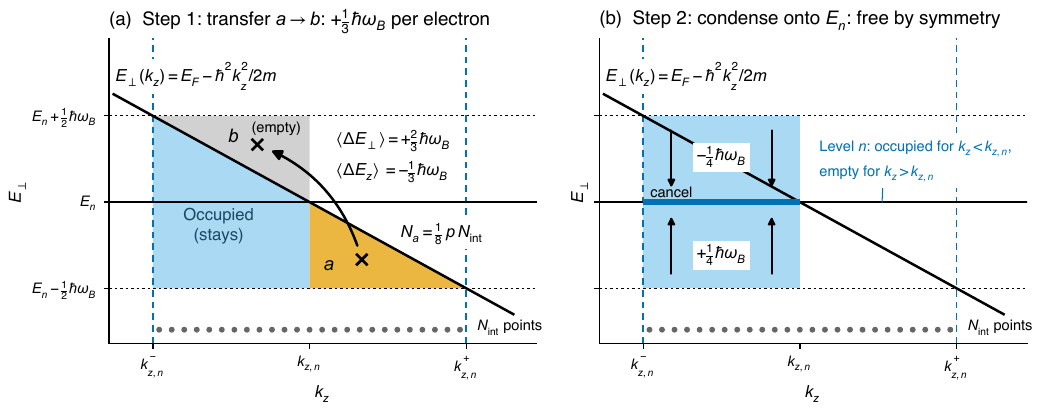}}
\caption{\label{fig:triangle}%
The triangle construction in one tiling interval (the interval marked in Fig.~\ref{fig:cylinders}(b), magnified; the boundary is locally linear).
(a)~Step 1: the occupied triangle $a$ (yellow), which contains
$N_a=\tfrac18 pN_{\rm int}$ electrons, is transferred to the congruent
empty triangle $b$. Crosses mark the centroids; the transfer costs
$+\tfrac23\hw$ in $\Ep$ and returns $\tfrac13\hw$ in $E_z$
[Eqs.~\eqref{eq:dEperp} and \eqref{eq:dEz}], a net $+\tfrac13\hw$ per
electron.
(b)~Step 2: the strip, now symmetric about $E_n$ for
$\kz<k_{z,n}$, condenses onto the level for free; the level ends up
occupied only for $\kz<k_{z,n}$, i.e., inside the Fermi surface.}
\end{figure*}

Consider free electrons in a box $V=L_x L_y L_z=A L_z$ with a uniform
field $\bm{B}=B\hat{z}$. The single-particle spectrum is
\begin{equation}
E_{n,\kz} \;=\; \hw\Bigl(n+\tfrac{1}{2}\Bigr) + \frac{\hbar^2 \kz^2}{2m},
\qquad \omega_B=\frac{eB}{m},
\label{eq:spectrum}
\end{equation}
with $n=0,1,2,\dots$ and $\kz=2\pi j/L_z$, $j$ integer. Each level carries the
degeneracy (spin included)
\begin{equation}
p \;=\; \frac{eBA}{\pi\hbar} \;=\; 2\,\frac{\Phi_B}{\Phi_0},
\label{eq:degeneracy}
\end{equation}
where $\Phi_B=BA$ and $\Phi_0=h/e$. The factor of two from spin
degeneracy is included in $p$, but Zeeman splitting is omitted from
Eq.~\eqref{eq:spectrum}: it modifies the dHvA amplitude and can produce
spin zeros, without changing the orbital counting used here. In $k$ space the states occupy
concentric \emph{Landau cylinders}, coaxial with $\bm{B}$, of radii
$k_{\perp,n}=\sqrt{2m E_n}/\hbar$, where $E_n\equiv\hw(n+\tfrac12)$ is the
perpendicular level energy [Fig.~\ref{fig:cylinders}(a)]. At $T=0$ and fixed electron number, the
occupied portion of each cylinder is the segment inside the Fermi sphere.

The entire argument of this paper takes place in a two-dimensional
projection of this picture [Fig.~\ref{fig:cylinders}(b)]: plot, against
$\kz$, the perpendicular energy $\Ep$. Landau cylinders become horizontal
lines at $E_n$, spaced by $\hw$. A state is occupied when its total energy $E_n+\hbar^2\kz^2/2m$ lies below $E_F$, so the occupation boundary is the parabola
\begin{equation}
\Ep(\kz) \;=\; E_F-\frac{\hbar^2 \kz^2}{2m},
\label{eq:boundary}
\end{equation}
and the occupied states at a given $\kz$ are the lines below it.
Energy and cross-sectional area are interchangeable on this plot: the
slice of the Fermi sphere at height $\kz$ is a disk of area
$S(\kz)=\pi k_\perp^2=2\pi m\,\Ep(\kz)/\hbar^2$, proportional to the
boundary energy, so each point of the parabola represents one cross
section. In particular the apex ($\kz=0$, $\Ep=E_F$) represents the equatorial (extremal) cross section, of area $S_0=\pi k_F^2$.
Figure~\ref{fig:cylinders}(b) anticipates the result of the next two
sections: the magnetic response comes entirely from the states within $\sim\hw$ of the occupation boundary; the interior is inert. The constant part comes from the flanks, and the oscillatory part from
the apex region alone.

For $E_F=5\;$eV and
$B=1\;$T, $\hw\approx 0.12\;$meV, so the number of occupied cylinders is
$N_L \approx E_F/\hw \approx 4\times10^4$, and
$p/A\approx 4.8\times10^{10}\;\text{cm}^{-2}$. All figures exaggerate
$\hw/E_F$ enormously for clarity.

\section{The tiling and the triangle construction}
\label{sec:triangle}

\subsection{Tiling the $\kz$ axis}

Group the energy axis into strips of width $\hw$ centered on the Landau
levels, $[E_n-\hw/2,\,E_n+\hw/2)$; in three dimensions this shell is the
magnetic tube of Ref.~\onlinecite{Nikolaev2018} (Appendix~\ref{app:nikolaev}), a
shell of thickness $\hw$ around one cylinder rather than the Landau tube
of Ref.~\onlinecite{AshcroftMermin}, which is the cylinder itself. When the field is
switched on, the zero-field states of strip $n$ (at fixed $\kz$) condense
onto the level $E_n$ (a regrouping of single-particle states; no
collective effect is implied): the strip contains
$p$ states per $\kz$ point both before and after, because the
two-dimensional density of states per unit area, $m/\pi\hbar^2$ including
spin, times $\hw$ equals exactly $p/A$.

For a strip that lies entirely below the occupation boundary, this
condensation costs nothing: the filled strip is symmetric about its
level, so the collapse raises the lower half by $\hw/4$ on average and
lowers the upper half by the same amount. The interior of the Fermi
sphere is therefore magnetically inert, as anticipated in Fig.~\ref{fig:cylinders}(b), and the entire energy change comes from the one partially filled strip per $\kz$ that the
boundary crosses.

As $|\kz|$ increases from $0$, the boundary parabola \eqref{eq:boundary}
descends monotonically through the strips, one after another. The $\kz$
axis is thereby partitioned, \emph{tiled}, into consecutive intervals
$[k_{z,n}^-,k_{z,n}^+]$, one per strip crossing, each containing
$N_{\rm int}\gg1$ discrete $\kz$ points [Fig.~\ref{fig:cylinders}(b)]. There
are $\approx 2N_L\sim10^5$ such intervals ($\pm\kz$), and every occupied
$\kz$ point belongs to exactly one of them. Within a single interval the
parabola may be linearized (on the far flanks its slope changes between neighboring intervals only
by a fraction of order $\hw/E_F$ of itself; the change grows toward the
apex), \emph{except} near the apex, a failure we turn
to in Sec.~\ref{sec:dhva}.

\subsection{One interval: transfer and condensation}

Figure~\ref{fig:triangle} magnifies one interval. The boundary crosses the
upper strip edge at $k_{z,n}^-$, the level $E_n$ at $k_{z,n}$, and the
lower strip edge at $k_{z,n}^+$; by linearity, $k_{z,n}$ bisects the
interval. At $B=0$ the occupied strip states are those below the boundary
line: for $\kz<k_{z,n}$ the column is filled beyond the level (blue region
of Fig.~\ref{fig:triangle}), while for $\kz>k_{z,n}$ the occupied states
form the triangle $a$ and the states of triangle $b$ are empty.

The ground state at $B\neq0$ is built in two steps.

\emph{Step 1: transfer.} Move the electrons of triangle $a$ into triangle
$b$. The two triangles are congruent, so the books balance locally and
exactly: no reservoir, no averaging assumption. The energy cost per electron follows from the centroids. The centroid of $a$, the mean of its three vertices, sits at $E_n-\tfrac13\hw$; that of $b$ sits at $E_n+\tfrac13\hw$. Hence
\begin{equation}
\langle\Delta \Ep\rangle=+\tfrac{2}{3}\hw .
\label{eq:dEperp}
\end{equation}
Simultaneously each transferred electron moves toward smaller $|\kz|$: the
centroid shift is $(k_{z,n}^+-k_{z,n}^-)/3$, and since
$E_z=\hbar^2\kz^2/2m$
varies by exactly $\hw$ across the interval along the boundary
(where $\Ep+E_z=E_F$), the longitudinal energy drops by
\begin{equation}
\langle\Delta E_z\rangle=-\tfrac{1}{3}\hw .
\label{eq:dEz}
\end{equation}
The net cost is $+\tfrac13\hw$ per transferred electron. The number of
transferred electrons is the area of triangle $a$: one eighth of the strip
rectangle, i.e.,
\begin{equation}
N_a=\tfrac{1}{8}\,p\,N_{\rm int}.
\label{eq:Na}
\end{equation}

\emph{Step 2: condensation.} After the transfer, the occupied part of the
strip for $\kz\in[k_{z,n}^-,k_{z,n}]$ fills the strip symmetrically about
the level: exactly as for the interior strips, the condensation onto $E_n$
is free. For $\kz>k_{z,n}$ the strip is empty and the level unoccupied: the boundary~\eqref{eq:boundary} still governs occupation, so level $n$
ends up occupied only on the half interval $[k_{z,n}^-,k_{z,n}]$, exactly
the segment of the Landau cylinder inside the Fermi sphere.

The energy cost of the interval is therefore
\begin{equation}
\Delta U_{\rm int}
= \tfrac{1}{8}\,p\,N_{\rm int}\times\tfrac{1}{3}\hw
= \tfrac{1}{24}\,p\,N_{\rm int}\,\hw ,
\label{eq:interval}
\end{equation}
that is, $\tfrac{1}{24}p\hw$ per $\kz$ point, \emph{independent of which interval}. The origin of Landau diamagnetism is therefore not the condensation, which is free by symmetry, but the small minority of electrons near each crossing of the occupation boundary, $\tfrac18 pN_{\rm int}$ per interval, that must rearrange along $\kz$ at a net cost of $\tfrac13\hw$ each.

\subsection{Summing the tiling: the Landau susceptibility}

Because the intervals tile the $\kz$ axis, the total is simply
$\tfrac1{24}p\hw$ times the number of occupied $\kz$ points,
$N_{\kz}=L_z k_F/\pi$:
\begin{equation}
\Delta U=\frac{1}{24}\,p\,\hw\,N_{\kz}
=\frac{V k_F e^2 B^2}{24\pi^2 m},
\label{eq:DeltaU}
\end{equation}
using Eq.~\eqref{eq:degeneracy}. The magnetization per unit volume,
$M=-V^{-1}(\partial\Delta U/\partial B)_{N,V}=-k_F e^2 B/12\pi^2 m$, then
gives, with $\chi=\mu_0 M/B$ the zero-field linear response,
\begin{equation}
\chi_L=-\frac{\mu_0 e^2 k_F}{12\pi^2 m}
=-\frac{1}{3}\,\mu_0\mu_B^2\,g(E_F)
=-\frac{\chi_P}{3},
\label{eq:chiL}
\end{equation}
with $g(E_F)=mk_F/\pi^2\hbar^2$ the density of states per volume and
$\chi_P$ the Pauli susceptibility. This is Landau's
result \cite{Landau1930}, obtained here by elementary state counting.

For an isotropic parabolic band in a crystal, the cyclotron mass in Eqs.~\eqref{eq:spectrum} and
\eqref{eq:DeltaU} is the effective mass $m^*$, while the Bohr magneton in
$\chi_P$ contains the bare mass $m_0$; hence
$\chi_L=-\tfrac13(m_0/m^*)^2\chi_P$ and
\begin{equation}
\chi=\chi_P\Bigl[1-\tfrac{1}{3}\bigl(m_0/m^*\bigr)^2\Bigr].
\label{eq:total}
\end{equation}
Within this isotropic single-parabolic-band model the net response is
diamagnetic for $m^*<m_0/\sqrt{3}\approx0.58\,m_0$: a small orbital mass
favors diamagnetism. For a real crystal the orbital response cannot in
general be reduced to a single scalar mass, since the cyclotron mass, the
density-of-states mass, the $g$ factor, the anisotropy of the Fermi
surface and interband contributions all enter separately;
Eq.~\eqref{eq:total} should be read as an illustration rather than as a
formula for a multiband material.

The only geometric approximation in the interval construction is the
linearization of the boundary across a single interval: the congruence of
the triangles, the centroid values, and the tiling are exact. The
derivation additionally uses the thermodynamic and weak-field limits
collected in Table~\ref{tab:ledger}. Table~\ref{tab:ledger}
collects what is assumed and what is not, so that the scope of the result
is not in doubt.

\begin{table*}[t]
\caption{\label{tab:ledger}What the derivation assumes.}
\footnotesize
\begin{ruledtabular}
\begin{tabular}{@{}ll@{}}
Ingredient & Status\\
\colrule
Congruence of the triangles, centroids, tiling & Exact\\
Linearization across one interval & The only approximation; controlled by
$\hw/E_F$ on the flanks, fails near the apex (Sec.~\ref{sec:dhva})\\
Discrete $\kz\to$ continuum, $N_{\rm int}\gg1$ & Thermodynamic limit; for a
rigorous treatment see Ref.~\onlinecite{Angelescu1975}\\
$\hw/E_F\ll1$ & Assumed throughout; $\sim10^{-5}$ at $B=1$~T\\
Fixed $N$: $E_n=\mu(B)$ versus $E_n=E_F$ & Differ by $10^{-3}$ of a level
spacing (Sec.~\ref{sec:dhva})\\
Zeeman splitting & Omitted (Sec.~\ref{sec:setup})\\
Scattering, interactions, finite $T$ & Outside the scope of this work\\
\end{tabular}
\end{ruledtabular}
\end{table*} Moreover, the
approximation is very well satisfied in practice: already at
$B=1\;$T the tiling comprises $\sim10^5$ intervals, and weaker fields only make the tiling finer. Yet the approximation fails at any field strength near the apex, where the slope of the boundary vanishes and
linearization fails not by degree but outright. We now follow the
construction to that point.

\section{Where the construction fails: the dHvA effect}
\label{sec:dhva}

\subsection{The equatorial interval}

\begin{figure*}[t]
\makebox[\textwidth][c]{\includegraphics{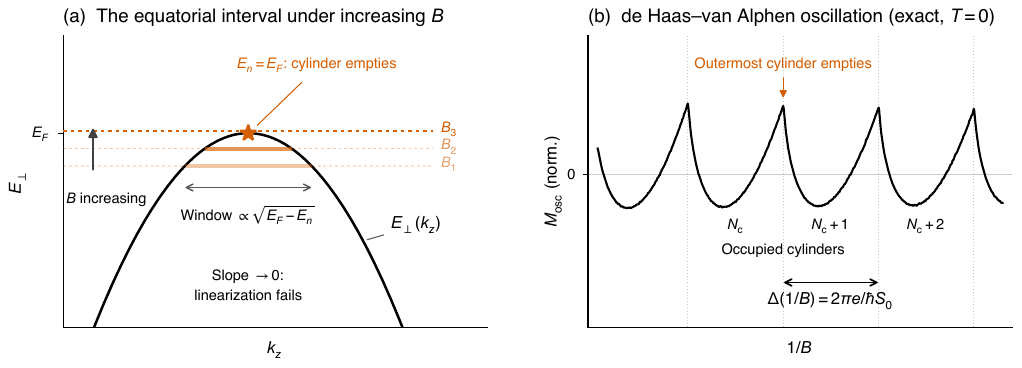}}
\caption{\label{fig:equatorial}%
(a)~The equatorial interval under increasing field. The topmost level
[the outermost cylinder of Fig.~\ref{fig:cylinders}(a)] rises with $B$
($B_1<B_2<B_3$); its occupied window \eqref{eq:window} shrinks like a square root and vanishes when
$E_n=E_F$ (star): one cylinder empties.
(b)~The resulting oscillation of the magnetization, computed at $T=0$ for
the free-electron gas in reduced units, by direct state counting as in
the Supplementary Information. Dotted
verticals mark the events of panel~(a); between events the number of
occupied cylinders $N_c$ is constant; the period is
Eq.~\eqref{eq:period}.}
\end{figure*}

Near the apex of the parabola, $\partial\Ep/\partial\kz\to0$: the interval
containing the extremal cross section cannot be linearized, the triangles
of Fig.~\ref{fig:triangle} are no longer congruent, and
Eq.~\eqref{eq:interval} fails there. The failure is graded rather than
abrupt. Counting intervals outward from the apex by $j=1,2,\dots$, the
boundary slope grows like $\sqrt{j}$, so it changes between neighbors by
$\sqrt{(j+1)/j}-1$, which is $41\%$ at $j=1$ and $22\%$ at $j=2$ but has
fallen to $10\%$ by $j=5$ and to $\sim1/2j$ thereafter. The geometric error is
therefore largest in the first two intervals and falls off rapidly
outward; the decomposition below shows that the summed residual is
already negligible beyond the first few. The population of the extremal interval is controlled
instead by the square-root window
\begin{equation}
|\kz| < k_{z,n} = \frac{\sqrt{2m(E_F-E_n)}}{\hbar}
\label{eq:window}
\end{equation}
of the topmost occupied level [Fig.~\ref{fig:equatorial}(a)], with
$k_{z,n}$ the crossing point of Sec.~\ref{sec:triangle}. As $B$
increases, the level ladder stretches, the window shrinks
like a square root, and when $E_n$ crosses $E_F$ the occupied
segment vanishes: \emph{the number of occupied Landau cylinders drops by
one}. At fixed electron number the event condition is strictly
$E_n=\mu(B)$ rather than $E_n=E_F$. The difference is immaterial here\footnote{See p.~273, footnote~17 of
Ref.~\onlinecite{AshcroftMermin}: the field dependence of the chemical potential at
fixed $N$ ``is a very small effect and can normally be ignored.''}: in
the numerics of the Supplementary Information $\mu(B)$ oscillates with a
peak-to-peak amplitude of $0.16\;\mu$eV, which is $1.4\times10^{-3}$ of a
level spacing and $3\times10^{-8}$ of $E_F$, so the period below is
unaffected to leading order in $\hw/E_F$. The equatorial contribution therefore resets at each event and varies smoothly between them, so the total energy, and with it the magnetization, oscillates.

The scale of the oscillation follows from the same picture: the extremal window contains a fraction $\sim\sqrt{\hw/E_F}$ of all $\kz$ points, about $5\times10^{-3}$ at $E_F=5$~eV and $B=1$~T, which is why the oscillatory part of the \emph{energy} amounts to only a fraction of a percent of the smooth term [Supplementary Information, Fig.~\ref{fig:numerics}(a)]; the magnetization, a derivative, is not suppressed by the same ratio.

The Fermi energy, by contrast, is pinned by the regular intervals, whose combined density of states is far
larger than that of the equatorial window; a cylinder crossing the Fermi surface at some generic
$\kz$ produces no leading singularity, being absorbed into the smooth
$1/24$ bookkeeping of
Sec.~\ref{sec:triangle}. The observable events are equatorial only. This
is the three-dimensional mechanism that the two-dimensional sawtooth
picture cannot supply: there, oscillations occur because the Fermi level
jumps between Landau levels; here it does not, and the oscillation records the birth and death
of cylinder segments at the extremal cross section.

What about the other intervals? The boundary is not exactly
straight across an interval, so each contributes a small residual beyond
Eq.~\eqref{eq:interval}, and that residual sweeps with $B$. But the
residuals largely cancel. The magnitude of a residual varies slowly, but
its sign is set by how the level ladder is registered against the
boundary, that is, by the number of level spacings that fit beneath the
boundary at $\kz$,
$\Ep(\kz)/\hw=\hbar S(\kz)/2\pi eB$. By the definition of the tiling
this count advances by exactly one across each interval, so the residual
runs through a full cycle of sign and averages away to leading order,
when the amplitude and the rate of advance vary little across it.
What survives is governed by the slow change of the sweep rate, cancels
in turn when summed along $\kz$, and leaves a net contribution only
where the sweep rate vanishes: at the extremum. Near the extremum the phase varies
quadratically with $\kz$, and summing the contributions there yields the
Fresnel pattern of Pippard's Cornu-spiral construction \cite{Pippard1960}.
The triangle transfer thus fixes the leading smooth contribution of the
regular intervals, and the stationary-phase argument explains why their
residuals add up only near the extremum.

Direct state counting settles how far the oscillatory anomaly extends. We decompose the
oscillatory residual bin by bin, using the fixed-$\mu$ bookkeeping defined
in the Supplementary Information, at $E_F=5\;$eV and
$B\approx1\;$T. The decomposition combines the symmetry-related intervals
at $\pm\kz$ into a single bin, so the $\sim10^5$ intervals of
Sec.~\ref{sec:triangle} appear as $43\,191$ bins at these fields. The
extremal bin and its first neighbor reproduce the residual to within
$1.2\%$ of its peak amplitude; taking the five nearest the apex, all
$43\,186$ remaining bins together contribute less than $0.2\%$ of that
peak at any field in the period [Supplementary Information,
Fig.~\ref{fig:numerics}(c)]. The rule that extremal cross sections dominate is, in this
precise sense, a statement about a handful of bins out of forty
thousand. This is the content, in the
present language, of the stationary-phase result of Lifshitz and
Kosevich \cite{LifshitzKosevich1956}.

\subsection{The period}

The vanishing condition $E_n=\hw(n+\tfrac12)=E_F$ states that the
cylinder's cross-sectional area,
$S_n=\pi k_{\perp,n}^2=(n+\tfrac12)\,2\pi eB/\hbar$, sweeps through the
extremal area $S_0=\pi k_F^2$. Successive events at fields $B_n$ therefore
satisfy $S_0=(n+\tfrac12)\,2\pi e B_n/\hbar$, i.e.,
\begin{equation}
\Delta\!\left(\frac{1}{B}\right)
=\frac{1}{B_{n}}-\frac{1}{B_{n-1}}
=\frac{2\pi e}{\hbar S_0},
\label{eq:period}
\end{equation}
the Onsager relation \cite{Onsager1952}, obtained here without
semiclassical quantization. Figure~\ref{fig:equatorial}(b) shows the oscillatory magnetization computed exactly at $T=0$ (see the Supplementary Information for the method): sharp maxima at the cylinder-emptying
events, the characteristic rounded minima in between, and the period
\eqref{eq:period}.

\section{Summary and discussion}
\label{sec:discussion}

One construction has carried the whole argument. The occupation
boundary descends through one Landau strip after another; each crossing
marks off one interval of the $\kz$ axis, and the intervals tile it.
Within each interval the energy cost of the field is a transfer of
states between two congruent triangles. The tiling gives the Landau
susceptibility. The narrow neighborhood where the tiling fails gives
everything else: the dHvA oscillation, its Onsager period, and the
extremal-orbit rule, all exhibited explicitly for the free-electron
sphere.

The regular-versus-extremal distinction also indicates how the picture
extends qualitatively beyond the Fermi sphere. For a general Fermi
surface the relevant profile is the cross-sectional area $S_F(\kz)$:
both its maxima (bellies) and minima (necks) are stationary cross
sections and therefore generate their own Onsager oscillations
\cite{LifshitzKosevich1956,Shoenberg}. Beyond the sphere the correspondence stays qualitative: the
$\tfrac1{24}$ coefficient rests on the equal level spacing and constant
density of states of the free-electron gas.

The construction refines a line of earlier arguments. Pippard's slice
picture \cite{Pippard1960} is the closest ancestor in the elementary
line: we replace his per-slice particle reservoir (and the attendant
$\tfrac14$ and $\tfrac12$ factors) by a closed-system transfer between
congruent triangles, and his Cornu-spiral summation reappears as the
interval-by-interval cancellation of Sec.~\ref{sec:dhva}. Dupr\'e's
uniform average over the values $x_i$ \cite{Dupre1981} is realized here
as the deterministic sweep of the boundary across one linearized
interval; Appendix~\ref{app:dupre} makes the correspondence, and the
difference in the longitudinal bookkeeping, explicit. The
density-of-states argument of Ashcroft and Mermin \cite{AshcroftMermin}
explains the singularity structure but not the energetics, which the
present construction supplies for both the steady and the oscillatory
response. For the free-electron gas, Nikolaev's magnetic tubes
\cite{Nikolaev2018} are the same shells counted in $k$ space;
Appendix~\ref{app:nikolaev} compares the two bookkeeping schemes.

Two questions were raised at the outset. \emph{Why is the electron gas
diamagnetic?} Because near each crossing of the occupation boundary, one
eighth of a Landau strip's electrons must move outward in energy by
$\tfrac23\hw$ while recovering only $\tfrac13\hw$ along the field.
\emph{Why do extremal orbits dominate the oscillations?} Because everywhere else the same bookkeeping is smooth to leading order; only where the occupation boundary is flat can a level's occupied piece close up, and the magnetization develop a cusp.

\appendix

\section{Relation to Dupr\'e's averaging}
\label{app:dupre}

Dupr\'e \cite{Dupre1981} defines $x_i=E_{\max}(k_z^i)-E_i$ for each
plane $k_z^i$, where $E_{\max}$ is the largest perpendicular energy the
Fermi surface allows in that plane and $E_i$ the corresponding Landau
level, and assumes the values of $x_i$ to be uniformly distributed on
$[-\hw/2,\hw/2]$. In our notation $E_{\max}(k_z^i)=\Ep(k_z^i)$ and
$E_i=E_n$, so his discrete variable corresponds to
$x(\kz)=\Ep(\kz)-E_n$.
In Fig.~\ref{fig:triangle}, $x$ is the height of the boundary above the
level: by the linearization of Sec.~\ref{sec:triangle}, $x$ sweeps
\emph{linearly and deterministically} from $+\hw/2$ to $-\hw/2$ as
$|\kz|$ increases across a regular interval. The uniform average over $x$ is
therefore realized, within the local linearization and the continuum
limit $N_{\rm int}\gg1$, as an interval average, and Dupr\'e's
moments $\langle x\rangle=0$, $\langle x^2\rangle=\hbar^2\omega_B^2/12$
are recovered as interval identities. His global bookkeeping of the
redistribution, which invokes equal statistical weights of positive and
negative $x_i$ so that excess and deficit populations balance after
summing over the planes, is replaced by a transfer between congruent
triangles that conserves the electron number within each linearized
interval.

The two calculations agree on the leading smooth coefficient but divide
the cost differently. Dupr\'e argues, from the assumed randomness of the $x_i$,
that the energy of the motion along the field is unchanged. Within an
interval, however, $x$ and $|\kz|$ are deterministically anticorrelated,
so the transferred electrons move systematically toward smaller $|\kz|$:
the longitudinal energy falls by $\tfrac13\hw$ each
[Eq.~\eqref{eq:dEz}]. Averaged over an interval, which contains
$N_a=\tfrac18 pN_{\rm int}$ transferred electrons, the two channels
contribute $+p\hw/12$ and $-p\hw/24$ per $\kz$ point, leaving the same net
$p\hw/24$ that Dupr\'e obtains.

\section{Relation to Nikolaev's magnetic tubes}
\label{app:nikolaev}

For the free-electron gas, Nikolaev \cite{Nikolaev2018} and the present
construction count the same states, but in different coordinates. He
counts in $k$ space, where the unit is a magnetic tube: the shell between
the auxiliary orbits of areas $A_n^{\rm aux}=2\pi eBn/\hbar$ and
$A_{n+1}^{\rm aux}$ [his Eqs.~(8) and (9)], with the $n$th Landau level in
the middle. Its cross sections form a chain of triangles in the
$(k_\perp,\kz)$ plane. Because the unit is built in $k$ space, the
construction is not tied to free electrons: within a weak-field
semiclassical treatment it extends to any simply connected Fermi surface,
the tube boundary then fixed by $A_n^{\rm aux}=2\pi eB(n+\delta)/\hbar$
with a small $\delta\sim B$. We count in the $(\kz,\Ep)$ plane instead. There the two-dimensional
density of states is constant, so equal areas hold equal numbers of
states. Only in these coordinates are the two transfer triangles of
Fig.~\ref{fig:triangle} congruent, and only there does the counting reduce
to centroid arithmetic.

The bookkeeping differs accordingly. Nikolaev works with the average
energy of all the active states in a tube, $\tfrac12 p$ of them per $\kz$
point. Before the field their transverse and longitudinal energy together
lie $\tfrac23\hw$ above his reference level on average; after it they
carry $\tfrac12\hw$ transversely and $\tfrac14\hw$ along the field, on
the same references [his Eqs.~(23)--(25)]. These are comparable
numbers, and the cost per state is their small difference,
$\tfrac12+\tfrac14-\tfrac23=\tfrac1{12}$, which returns the
$\tfrac1{24}p\hw$ per $\kz$ point of Eq.~\eqref{eq:interval}. Our
bookkeeping builds the same cancellation into the geometry instead: the
symmetric part cancels exactly, and the net $\tfrac1{24}$ is represented
by the transfer of $\tfrac18 p$ electrons at $+\tfrac13\hw$ each. Because
the cost sits in this minority, the construction connects to Dupr\'e's
averaging (Appendix~\ref{app:dupre}) and lets the oscillation be resolved
bin by bin in Sec.~\ref{sec:dhva}. In exchange, the tube formulation is
more general. It asks only for the shape of the Fermi surface and the energy
gradient on it, so it extends to real, anisotropic metals and \emph{ab
initio} input. What it gives up is the elementary arithmetic: the
coefficient now sits inside a Fermi-surface integral over the cyclotron
mass $m^{*}(\kz)$, which must come from the band structure. The two
formulations trade scope against simplicity, and the present one makes
the opposite choice.

\bibliography{refs}

\onecolumngrid
\clearpage
\setcounter{figure}{0}
\setcounter{equation}{0}
\renewcommand{\thefigure}{S\arabic{figure}}
\renewcommand{\theequation}{S\arabic{equation}}
\renewcommand{\theHfigure}{S\arabic{figure}}
\renewcommand{\theHequation}{S\arabic{equation}}

\vspace*{6pt}
{\Large\sffamily\bfseries\raggedright Supplementary Information for ``Landau diamagnetism and the
de Haas--van Alphen effect from a single geometric construction''\par}
\vspace{3pt}
{\raggedright\leftskip=0.5in\sffamily Sung-Hoon Lee\par}
\vspace{3pt}
{\raggedright\leftskip=0.5in\small\itshape Department of Applied Physics, Kyung Hee University,
Yongin, Republic of Korea\par}
\vspace{18pt}
\twocolumngrid

Notation follows the main text. We verify the three central claims of
the main text by direct occupation
of the spectrum of Eq.~\eqref{eq:spectrum} of the main text at $T=0$ and fixed $N$, at
the realistic parameters of Sec.~\ref{sec:setup} of the main text: $E_F=5$~eV, the bare electron mass,
spin degeneracy included, and fields near $B=1$~T, where
approximately $43\,190$ levels are occupied (the count changes by one at
each dHvA event). Two $\kz$ grids are used:
$L_z=0.52$~mm for the field scan (at least 11 $\kz$ points in the
narrowest tiling interval) and $L_z=1.6$~mm for the interval
decomposition (at least 33 points); since a slice with $\nu$ occupied
levels has the closed-form energy $p\,[\hw\,\nu^2/2+E_z \nu]$, no
state-by-state enumeration is needed. The Fermi level at fixed $N$ is
located by bisection (24 steps, resolving $\mu$ to a few times
$10^{-7}\,\hw$), with fractional occupation of the topmost level;
$\Delta U(B)=U(B)-U(0)$, with $U(0)$ evaluated on the same $\kz$ grid
with a continuum perpendicular spectrum. The same bisection gives the
field dependence of the chemical potential: over one dHvA period near
$B=1$~T, $\mu(B)-E_F^0$ oscillates with a peak-to-peak amplitude of
$0.16\;\mu$eV, i.e.\ $1.4\times10^{-3}\,\hw$, the value quoted in
Sec.~\ref{sec:dhva} of the main text.

\emph{(i) The $1/24$ coefficient.} A least-squares fit of $cB^2$ to the
full $\Delta U(B)$ over the first seven whole dHvA periods above
$1/B=1\;\text{T}^{-1}$, a window that strongly suppresses the
oscillatory bias in the fitted smooth coefficient, gives $\bigl|c/\bigl[Vk_Fe^2/24\pi^2m\bigr]-1\bigr| < 2\times10^{-6}$: Eq.~\eqref{eq:DeltaU} of
the main text is reproduced to the accuracy of the fit, and the oscillation
about the smooth law stays within $0.25\%$ at these fields
[Fig.~\ref{fig:numerics}(a)].

\emph{(ii) The period.} The oscillatory part of
$M=-\partial\Delta U/\partial B$ has maxima at the cylinder-emptying
events. On a scan grid incommensurate with the predicted period, with
event positions refined by local quadratic interpolation, the six
adjacent-event spacings equal the Onsager period
$2\pi e/\hbar S_0=2.3154\times10^{-5}\;\text{T}^{-1}$ to within
$0.3\%$ [Fig.~\ref{fig:numerics}(b)].

\emph{(iii) Localization of the oscillation.} We decompose the
oscillatory part of the energy over the tiling intervals. Slice rounding
leaves a small number mismatch $\Delta N_{\rm int}$ per interval, so the
additive quantity is not $\Delta U_{\rm int}$ but the transfer-credited
combination
\begin{equation}
\Omega_{\rm int}\;=\;\Delta U_{\rm int}-E_F^0\,\Delta N_{\rm int},
\label{eq:omegaint}
\end{equation}
in which every exchanged electron is charged at $E_F^0$, the energy it
carries on the occupation boundary. The decomposition is indexed by
$|\kz|$, so each bin combines the symmetry-related tiling intervals at
$+\kz$ and $-\kz$; the $43\,191$ bins resolved here therefore correspond
to about twice as many connected intervals on the full $\kz$ axis. Equation~\eqref{eq:omegaint} is a
grand-potential difference at fixed $\mu=E_F^0$, used here as a
diagnostic; it is the numerical counterpart of the closed-system triangle
transfer, and without the credit an interval is not a closed system and
the decomposition is not additive. Summing
$\Omega_{\rm int}$ over all bins reproduces the corresponding fixed-$\mu$
grand-potential difference to $4\times10^{-16}$ of its magnitude, so the
decomposition is exhaustive. It serves as a diagnostic for the fixed-$N$
energy of (i) and (ii) and is not identified with it: at these parameters
the two differ by a few parts in $10^6$ of $\Delta U$.

\onecolumngrid
\vspace*{\fill}
\vspace{\floatsep}
\begin{figurehere}
\makebox[\textwidth][c]{\includegraphics{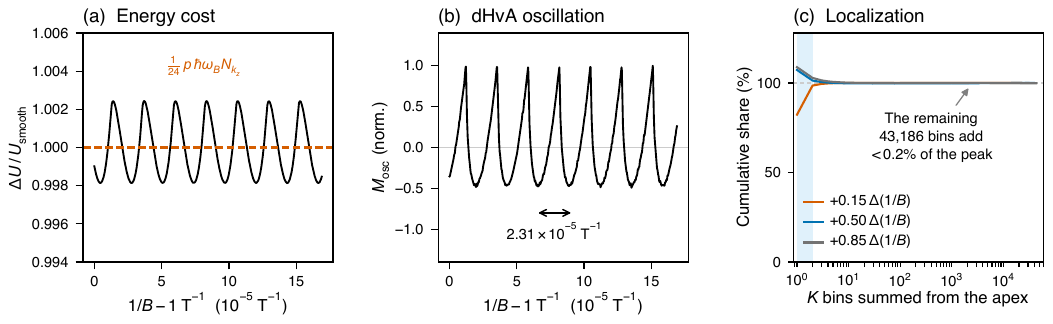}}
\caption{\label{fig:numerics}%
Exact $T=0$ numerics for the free-electron gas at $E_F=5$~eV, bare
electron mass, and fields near $B=1$~T.
(a)~$\Delta U$ divided by the smooth prediction of Eq.~\eqref{eq:DeltaU} of the main
text: the exact result oscillates about $1$ by less than $0.25\%$.
(b)~Oscillatory magnetization; the oscillation repeats with the Onsager
period $2\pi e/\hbar S_0=2.31\times10^{-5}\;\text{T}^{-1}$, marked
between adjacent minima. (c)~Cumulative signed share of the total oscillatory residual obtained by
summing the $K$ bins nearest the apex, at three fields spanning one dHvA
period. It converges within a few bins and is flat thereafter; beyond the
five nearest the apex the remaining $43\,186$ add less than $0.2\%$ of
the peak amplitude of the total residual.}
\end{figurehere}
\clearpage
\twocolumngrid

A statement about localization is a statement about a \emph{sum}: many
individually negligible residuals could still add up. We therefore
accumulate the residual $\Omega_{\rm int}-\tfrac1{24}pN_{\rm int}\hw$
outward from the apex [Fig.~\ref{fig:numerics}(c)]. Normalized to the peak amplitude of the total residual over the period,
the extremal bin and its first neighbor reproduce it to within
$1.2\%$; including the five bins nearest the apex, the summed
contribution of the remaining $43\,186$ stays below $0.2\%$ of that peak
at every field. The cumulative share is flat thereafter over four
decades. Individually those intervals
sit at the $\tfrac1{24}$ prediction, with a median of $1.0000$ and the
central $98\%$ within $\pm1.3\%$.

\end{document}